\documentclass{article}
\usepackage{bbm}
\usepackage{authblk}
\usepackage{a4wide}
\usepackage{dsfont}            % used in the commands \I, \Ww, \wW, \WW
\usepackage{amssymb,amsmath,amsfonts}
\usepackage{mathrsfs}
\usepackage{graphicx} 
\usepackage{leftidx}
\usepackage{xcolor}          % used in the commands \wW, \WW
\usepackage[utf8]{inputenc}% needed for Stratila's initial
\usepackage[all]{xy}
\usepackage{placeins}
\usepackage[version=4]{mhchem} 
\usepackage{textcomp}
\usepackage{siunitx}
\usepackage{pdfpages} 
\usepackage{url}
\usepackage{placeins}

\newcommand{\CC}{\mathbb{C}}
\newcommand{\I}{\mathds{1}}

\DeclareMathOperator*{\argmin}{arg\,min}

\begin{document} 
 

# Resolution Enhancement in Protein NMR Spectra by Deconvolution with Compressed Sensing Reconstruction

Krzysztof Kazimierczuk,[a] Paweł Kasprzak,[a,b] Panagiota S. Georgoulia,[c] Irena Matečko-Burmann,[d,e] Björn M. Burmann,[c,e] Linnéa Isaksson,[c] Emil Gustavsson,[f] Sebastian Westenhoff,[c] and Vladislav Yu. Orekhov*[c,f]

[a]   Centre of New Technologies, University of Warsaw, ul. Banacha 2C, 02-097 Warsaw, Poland
[b]   Faculty of Physics, University of Warsaw, Pasteura 5, 02-093 Warsaw, Poland
[c]   Department of Chemistry and Molecular Biology, University of Gothenburg, Box 465, Gothenburg 405 30, Sweden
[d]   Department of Psychiatry and Neurochemistry, University of Gothenburg, Gothenburg 405 30, Sweden
[e]   Wallenberg Centre for Molecular and Translational Medicine, University of Gothenburg, Gothenburg 405 30, Sweden
[f]   Swedish NMR Centre, University of Gothenburg, Box 465, 405 30, Gothenburg, Sweden.
      E-mail: vladislav.orekhov@nmr.gu.se

Supporting information for this article is given via a link at the end of the document.

**Abstract:** Multidimensional NMR spectroscopy is one of the basic tools for determining the structure of biomolecules. Unfortunately, the resolution of the spectra is often limited by inter-nuclear couplings. This limitation cannot be overcome by common ways of increasing resolution, i.e. non-uniform sampling (NUS) followed by compressed sensing (CS) reconstruction. In this paper, we show how to enrich CS processing with virtual decoupling leading to an increase in resolution, sensitivity, and overall quality of NUS reconstruction. A mathematical description of the decoupling by deconvolution approach explains the effects of noise, modulation of the sampling schedule, and reveals relation with the underlying assumption of the CS. The gain in resolution and sensitivity is demonstrated for the basic experiment used for protein backbone assignment 3D HNCA applied to two large protein systems: intrinsically disordered 441-residue Tau and a 509-residue globular bacteriophytochrome fragment.

Nuclear magnetic resonance (NMR) is among the main analytical techniques allowing atomic-level studies of proteins. The prerequisite step for most of protein NMR work is a resonance-specific spectral assignment, i.e. association of resonance frequencies with atoms in the protein amino acid chain.[1] The HNCA[2] is by far the most sensitive and thus often the only feasible triple resonance experiment that provides sequential connectivities between neighbouring protein residues. In principle, a sequence-specific resonance assignment could be obtained using the HNCA experiment alone. Unfortunately, low signal resolution relative to the dispersion of the protein $^{13}C^{\alpha}$ resonances results in massive ambiguity of the assignment even for relatively small proteins. For large systems with many amino acid residues as well as for intrinsically disordered proteins (IDP) characterized by the particularly low resonance dispersion, one has to rely on additional experiments at the expense of sensitivity loss, a significant increase of measurement time, and more complicated and tedious analyses.

Slow transverse relaxation of the $^{13}C^{\alpha}$ spins, which can be further decreased by deuteration, corresponds to the natural line-width of 5–8 Hz even for relatively large protein systems. Unfortunately, the practical resolution in the HNCA spectra is usually almost ten times worse. Two main factors limit the resolution in the HNCA: (i) a large number of time-increments in the $^{13}C^{\alpha}$ dimension needed in the 3D experiment to achieve the high resolution. This leads to too long measurement time that can be unaffordable because of short sample stability and/or limitation on the measurement time at an NMR instrument; (ii) homonuclear one-bond coupling between $^{13}C^{\alpha}$ and $^{13}C^{\beta}$ spins that produces a doublet with separation of approximately 35 Hz for every $^{13}C^{\alpha}$ signal and thus effectively broadens the spectral line. The former issue is well addressed by using fast pulsing[3] and non-uniform sampling (NUS) techniques.[4] A large number of methods for handling of the $^1J(C^{\alpha}–C^{\beta})$ coupling had been introduced over the last decades, including biochemical unlabelling of the β carbon atom to $^{12}C$,[5] constant time evolution,[6] band-selective homonuclear decoupling,[7] and IPAP decoupling.[8] However, broad practical use of these techniques is hindered due to the inherent compromises in sensitivity, extra demands on sample isotope labelling, inability to deal with serine and threonine residues, and/or significant spectral distortions and artefacts. A viable alternative to these experimental approaches is the virtual decoupling that is the post-acquisition deconvolution of the J-coupling at the signal processing stage.[9] The aim of this communication is to investigate the possibility of effective deconvolution in compressed sensing (CS) algorithms that are among the most powerful for the NUS spectra – to propose a method of selective deconvolution of individual spectral regions; and to demonstrate relation of the deconvolution to the cornerstone CS concept of sparseness with the resulting benefits for the effectiveness of CS.

The fundamental relation between the NMR signal $f(t)$ detected in the time domain and the spectrum $s$ is

$$f = Fs \qquad (1)$$

where $F$ is the measurement matrix composed of rows from the inverse Fourier transform matrix for every point in $f$. Thus, reconstruction of a spectrum from $f$ reduces to solving the inverse linear system in Eq. 1. For an undersampled (i.e. NUS) signal, the solution is not unique and additional constraints on the spectrum $s$ are usually imposed. For example, using generalized Tikhonov regularization, the spectrum can be obtained as:

$$s = arg\,min_x\bigl(\|Fx - f\|_Q^2 + \|x\|_D^2\bigr) \qquad (2)$$

where, for a vector $y$ and matrix $G$, $\|y\|_G^2$ denotes the weighted norm square $y^\dagger G y$ with $y^\dagger$ denoting the conjugate transpose of $y$; $Q = \sigma^{-2} I$ is the inverse covariance matrix of noise in $f$, which is multiple of the identity matrix $I$ and $\sigma$ is the standard deviation of the noise; $D$ is a diagonal matrix including weightings of the spectrum points and the Tikhonov regularization term. As will be shown below, $Q$ is useful when dealing with the $^1J(C^\alpha–C^\beta)$ coupling, while matrix $D$ is the essential element of the Iterative Reweighted Least Squares (IRLS), one of the most popular algorithms for compressed sensing reconstruction of the NUS spectra (see Supporting information)[10].

Assuming the same value of active $^1J(C^\alpha–C^\beta)$ coupling for all signals, the measured in experiment $^{13}C^\alpha$ signal $f$ and the signal without the J-coupling $\tilde{f}$ are related as:

$$f = C\tilde{f} \text{ and } \tilde{f} = C^{-1} f \quad (3)$$

where $C$ is a diagonal matrix with elements $cos(\pi J t)$ for every time point $t$ in $f$. If points in the measured signal $f$ are corrupted by noise with inverse covariance matrix $Q = \sigma^{-2} I$, the noise in $\tilde{f}$ has the inverse covariance matrix $\tilde{Q} = \sigma^{-2}(CC)$. Then, the decoupled spectrum is

$$\tilde{s} = arg\,min_x \left( \|Fx - \tilde{f}\|_{\tilde{Q}}^2 + \|x\|_D^2 \right) \quad (4a)$$

or equivalently (see Supporting information),

$$\tilde{s} = arg\,min_x \left( \|CFx - f\|_Q^2 + \|x\|_D^2 \right) \quad (4b)$$

The last equation shows that the post-acquisition deconvolution can be achieved in IRLS and any other algorithm based on equation akin to Eq 2, e.g. Maximum Entropy[9a-d] and Multi-Dimensional Decomposition (MDD),[4c] by using measurement matrix $CF$ instead of $F$. Finally, we note that the deconvoluted spectrum contains half of the peaks relative to the undecoupled spectrum. Thus, it is *sparser*, and in accordance with the theory of compressed sensing,[11] it requires nearly half of the measured data points for successful reconstruction. This means that the virtual decoupling not only enhances spectral resolution but also provides conditions for higher quality CS reconstruction (see theory in SI).

Use of the deconvolution for the HNCA experiment is based on the assumption that $^1J(C^\alpha–C^\beta)$ coupling constants are nearly the same for all residues in the protein. The variation of the coupling values ±2.5 Hz[12] is lower than the line width determined by the transverse relaxation of $^{13}C^\alpha$ spins and, thus, does not pose a problem for the reconstruction (see Supporting information). However, signals (singlets) from Gly residues that do not have $C^\beta$ atoms, have no sparse representation in the columns of measurement matrix $CF$. Figure 1 illustrates that this not only corrupts the Gly peaks in the deconvoluted spectrum but also affects other signals and reduces the overall quality of the reconstruction. To tackle this, we suggest a procedure of deconvolution-IRLS (D-IRLS) with the Gly-region selection as outlined in Figure 1 (more details are found in the Supporting information). We start with reconstructing the full undecoupled spectrum using matrix $F$. Because the $^{13}C^\alpha$ atoms usually have distinctly different chemical shifts with values lower than 45 ppm, we can subtract the well-reproduced signals in the Gly region from the original time-domain signal $f$, which is then used to reconstruct the spectrum with all signals except for Gly using Eq.

4b with the measurement matrix $CF$. Finally, signals of Gly and other residues are combined into the full decoupled spectrum in the frequency domain.

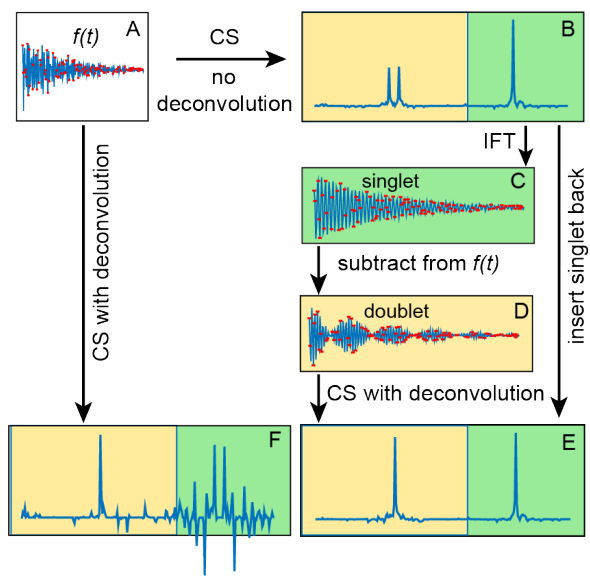

**Figure 1.** Processing of a spectrum with region-selective deconvolution. **A** – measured time-domain signal that contains both a singlet and a doublet. IRLS reconstructions of A with and without deconvolution produce spectra **F** and **B**, respectively. **C** - the singlet (green) part of the spectrum B is converted back to the time-domain using inverse Fourier transform (IFT). **D** – the original signal A after subtraction of D. **E** – the result of the region-selective deconvolution, i.e. combination of IRLS processing of D (yellow) and the green part of B. Quality of both singlet and decoupled doublet signals in E is better than in F.

Selection of the NUS acquisition schedule has a profound effect on the reconstruction quality. As $f$ is multiplied by $C^{-1}$ in Eq. 3, the noise is amplified the most for the points in $\tilde{f}$ at times, where $cos(\pi J t)$ function has small values (i.e. near $t = k/(2J), k = 1,3,5$). In the weighted least squares method used to derive Eq. 4, these points are used with low weights and thus contain relatively low information value. In the NUS schedule, it is logical to avoid these points and instead invest spectrometer time into more informative measurements. We used the signal amplitude matched NUS schedule with the sampling density corresponding to $|cos(\pi J t)|$ and rejecting points with probability less than 0.2[13] (Supporting Figure S1). Additionally, the schedule was in all cases relaxation-matched.

We demonstrate the new D-IRLS procedure using examples of two representative systems: intrinsically disordered human 441-residue Tau protein (the longest hTau40 isoform)[14] and the monomeric variant of the 509-residue globular photosensory module PAS-GAF-PHY of *Deinococcus radiodurans* phytochrome (*Dr*BphP$_{PSM}$).[15] For each protein, Figure 2 shows the traditional low-resolution 3D HNCA spectrum superimposed with the resolution-enhanced spectrum obtained using D-IRLS with Gly-region selection. For *Dr*BphP$_{PSM}$ the two experiments were reconstructed using nearly the same number of NUS points corresponding to the same measurement time; for Tau, the low resolution experiment was around two times shorter. In the shown examples, the dramatically improved resolution of the D-IRLS spectrum allows us to observe sequential connectivities that are ambiguous in the traditional spectrum.

Figures 2, S2, and S3 demonstrate that, in addition to the enhanced resolution, the D-IRLS spectra show higher or similar sensitivity in comparison to both the traditional low resolution and non-deconvoluted spectra. The peak connecting A87 and A88 in $Dr$BphP$_{PSM}$ spectra (Figure 2B) provides a specific example of this. It is clearly seen in the 1D cross-sections in the D-IRLS spectrum. In the traditional experiment, the weak peak is completely masked by the slope of a stronger peak. In the non-deconvoluted spectrum, only one of the doublet components is present, which gives a completely wrong idea of the peak position.

reconstructed with Gly-region selective D-IRLS. While the improved resolution in the spectra is anticipated from the deconvolution, the remarkable sensitivity of the D-IRLS spectrum can be explained by the increased sparsity favourable for the NUS reconstruction.

In order to extensively test the proposed D-IRLS method, we conducted simulations using synthetic peaks added to the 3D HNCA signal of Tau. Adding the simulated components with known positions and intensities to the time domain signal makes it possible to define the precision of the corresponding peak parameters derived from the reconstructed spectrum.[16] A detailed description of the simulations can be found in the Supporting information. The results shown in Supporting Figure S3 confirm that the peak intensities and positions are much more accurate when the D-IRLS deconvolution is augmented with the Gly-region selective procedure (Figure 1E).

Notably, at low sampling levels (250 and 400 NUS points) the number of detected peaks from the lowest intensity fraction of the injected peaks is significantly larger in the deconvoluted spectrum in comparison to the IRLS without the deconvolution. This is fully in line with the theoretical consideration that the deconvoluted spectrum is much more sparse and thus can be successfully reconstructed with fewer measured points. In the wide range of NUS levels, cosine J-modulated sampling scheme provides spectra with comparable or somewhat better accuracy of the peak positions and intensities than the schedules matched to the exponential relaxation decay only. However, the main practical problem with the latter scheme is the necessity to increase and carefully adjust the Tikhonov regularization parameter λ in the IRLS algorithm, whereas the cosine-modulated sampling is much less demanding in this respect and thus more robust. It is also worth noting that precision of the peak positions derived from the non-deconvoluted spectrum is somewhat better than in its decoupled counterpart provided that both components of the doublet are detected and resolved from other peaks.

In conclusion, we proposed an efficient CS method to improve the resolution, sensitivity and quality of NUS reconstruction using virtual decoupling at the processing stage. We presented a complete mathematical description of the problem in terms of the generalized Tikhonov regularization formalism. We also showed that removing singlets from the spectrum before decoupling significantly improves results. The method was demonstrated on the 3D HNCA spectra of two large systems prototypical for the intrinsically disordered and globular proteins. The new CS virtual decoupling technique will enable the sequential signal assignment for many challenging proteins and will be useful for other types of NMR spectra in a variety of applications.

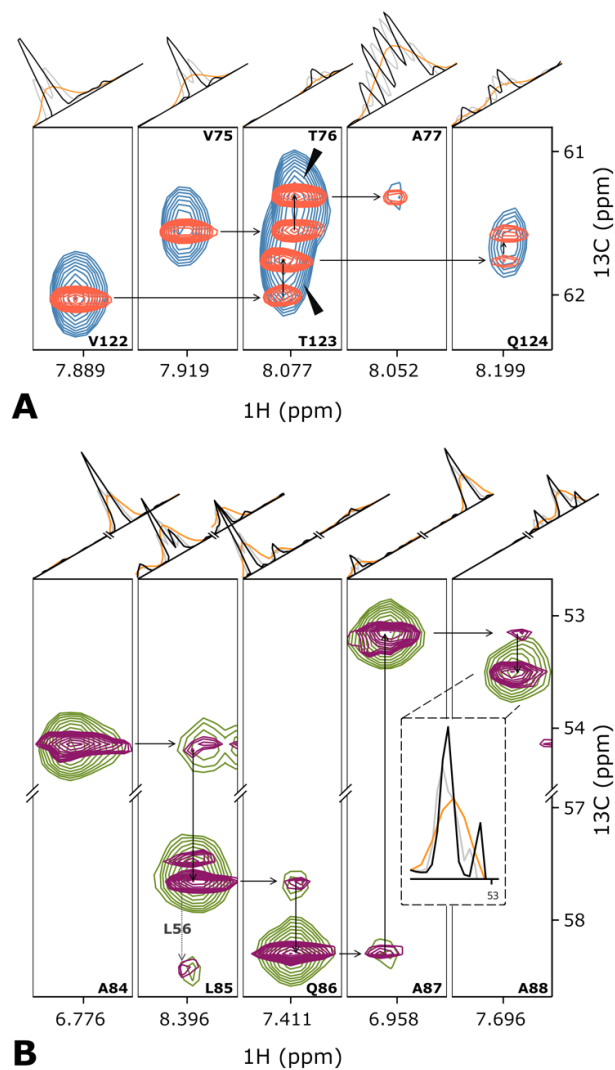

**Figure 2.** Several planes from the 3D HNCA spectra of **A** – Tau and **B** – $Dr$BphP$_{PSM}$ showing the assignment walk for selected residues. Overlaid blue (green) and red (purple) contour levels depict traditional low-resolution and high-resolution spectra of Tau ($Dr$BphP$_{PSM}$) protein. For the peak annotations, we use the previously published assignment.[14-15] The one-dimensional cross-sections above the spectra planes are taken (orange) the low-resolution, (black) high-resolution deconvoluted, and (grey) high-resolution non-deconvoluted spectra.

Figure S2 shows the $^{13}$C/$^{15}$N projections from the spectra of both studied proteins, which confirms the superior quality of the spectra

## Acknowledgements

## Supporting information

The Supporting information contains extended mathematical description of D-IRLS, sample preparation and experimental details, and results of simulations with injected peaks.

**Keywords:** biomolecular NMR • J-coupling • deconvolution • compressed sensing • non-uniform sampling

[1] M. Ikura, L. E. Kay, A. Bax, *Biochemistry* **1990**, *29*, 4659-4667.
[2] a) Z. Solyom, M. Schwarten, L. Geist, R. Konrat, D. Willbold, B. Brutscher, *J. Biomol. NMR* **2013**, *55*, 311-321; b) M. Salzmann, K. Pervushin, G. Wider, H. Senn, K. Wüthrich, *Proc. Natl. Acad. Sci. USA* **1998**, *95*, 13585-13590; c) L. E. Kay, M. Ikura, R. Tschudin, A. Bax, *J. Magn. Reson.* **1990**, *89*, 496-514.
[3] a) E. Lescop, P. Schanda, B. Brutscher, *J. Magn. Reson.* **2007**, *187*, 163-169; b) K. Pervushin, B. Vögeli, A. Eletsky, *J. Am. Chem. Soc.* **2002**, *124*, 12898-12902.
[4] a) J. C. J. Barna, E. D. Laue, M. R. Mayger, J. Skilling, S. J. P. Worrall, *J. Magn. Reson.* **1987**, *73*, 69-77; b) J. C. Hoch, A. S. Stern, M. Mobli, *Vol. 26*, John Wiley & Sons, Ltd, **2007**, pp. 1-6; c) V. Y. Orekhov, V. A. Jaravine, *Prog. Nucl. Magn. Reson. Spectrosc.* **2011**, *59*, 271-292; d) K. Kazimierczuk, V. Orekhov, *Magn. Reson. Chem.* **2015**, *53*, 921-926; e) M. Bostock, D. Nietlispach, *Concept Magn. Reson. A* **2017**, *46*, e21438; f) S. Robson, H. Arthanari, S. G. Hyberts, G. Wagner, in *Methods Enzymol., Vol. 614*, Elsevier, **2019**, pp. 263-291.
[5] a) D. M. LeMaster, D. M. Kushlan, *J. Am. Chem. Soc.* **1996**, *118*, 9255-9264; b) P. E. Coughlin, F. E. Anderson, E. J. Oliver, J. M. Brown, S. W. Homans, S. Pollak, J. W. Lustbader, *J. Am. Chem. Soc.* **1999**, *121*, 11871-11874; c) M. Kainosho, T. Torizawa, Y. Iwashita, T. Terauchi, A. M. Ono, P. Güntert, *Nature* **2006**, *440*, 52-57; d) P. Lundström, K. Teilum, T. Carstensen, I. Bezsonova, S. Wiesner, D. F. Hansen, T. L. Religa, M. Akke, L. E. Kay, *J. Biomol. NMR* **2007**, *38*, 199-212; e) K. Takeuchi, Z.-Y. J. Sun, G. Wagner, *J. Am. Chem. Soc.* **2008**, *130*, 17210-17211; f) S. A. Robson, K. Takeuchi, A. Boeszoermenyi, P. W. Coote, A. Dubey, S. Hyberts, G. Wagner, H. Arthanari, *Nat. Commun.* **2018**, 1-11.
[6] R. Powers, A. M. Gronenborn, G. M. Clore, A. Bax, *J. Magn. Reson.* **1991**, *94*, 209–213.
[7] a) H. Matsuo, E. Kupče, H. Li, G. Wagner, *J. Magn. Reson. B* **1996**, *113*, 91-96; b) P. W. Coote, S. A. Robson, A. Dubey, A. Boeszoermenyi, M. Zhao, G. Wagner, H. Arthanari, *Nat. Commun.* **2018**, 1-9.
[8] a) M. Ottiger, F. Delaglio, A. Bax, *J. Magn. Reson.* **1998**, *131*, 373-378; b) P. Andersson, J. Weigelt, G. Otting, *J. Biomol. NMR* **1998**, *12*, 435-441.
[9] a) A. A. Bothner-By, J. Dadok, *J. Magn. Reson.* **1987**, *72*, 540-543; b) M. A. Delsuc, G. C. Levy, *J. Magn. Reson.* **1988**, *76*, 306-315; c) Z. Serber, C. Richter, D. Moskau, J.-M. Böhlen, T. Gerfin, D. Marek, M. Häberli, L. Baselgia, F. Laukien, A. S. Stern, J. C. Hoch, V. Dötsch, *J. Am. Chem. Soc.* **2000**, *122*, 3554-3555; d) N. Shimba, A. S. Stern, C. S. Craik, J. C. Hoch, V. Dötsch, *J. Am. Chem. Soc.* **2003**, *125*, 2382-2383; e) R. Kerfah, O. Hamelin, J. Boisbouvier, D. Marion, *J. Biomol. NMR* **2015**, *63*, 389-402.
[10] a) E. J. Candes, M. B. Wakin, S. P. Boyd, *J. Fourier Anal. Appl.* **2008**, *14*, 877-905; b) K. Kazimierczuk, V. Y. Orekhov, *J. Magn. Reson.* **2012**, *223*, 1-10.
[11] a) S. Foucart, H. Rauhut, *Bull. Am. Math* **2017**, *54*, 151-165; b) E. J. Candès, J. Romberg, T. Tao, *IEEE Trans. Inf Theory* **2006**, *52*, 489-509.
[12] J. M. Schmidt, M. J. Howard, M. Maestre‐Martínez, C. S. Pérez, F. Löhr, *Magn. Reson. Chem.* **2009**, *47*, 16-30.
[13] V. Jaravine, I. Ibraghimov, V. Y. Orekhov, *Nat. Methods* **2006**, *3*, 605-607.
[14] R. L. Narayanan, U. H. Dürr, S. Bibow, J. Biernat, E. Mandelkow, M. Zweckstetter, *J. Am. Chem. Soc.* **2010**, *132*, 11906-11907.
[15] E. Gustavsson, L. Isaksson, C. Persson, M. Mayzel, U. Brath, L. Vrhovac, J. A. Ihalainen, B. G. Karlsson, V. Orekhov, S. Westenhoff, *Biophys. J.* **2020**, *118*, 415-421.
[16] A. D. Schuyler, J. C. Hoch, *NUScon, nonuniform sampling and reconstruction challenge in NMR spectroscopy.*, https://nuscon.org/, **2018**.

**Entry for the Table of Contents**

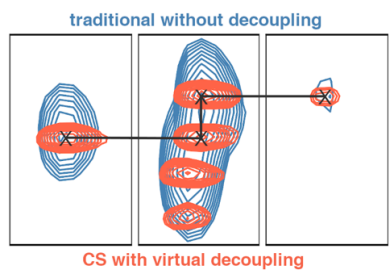

We introduce an algorithm of Compressed Sensing (CS) with virtual decoupling that increases resolution, sensitivity, and quality of NUS reconstruction of NMR spectra as exemplified by HNCA experiments for two large protein systems: intrinsically disordered 441-residue Tau and a 509-residue globular bacteriophytochrome fragment. Rigorous mathematical description of the algorithm in relation to the underlying assumption of the CS is presented.

 
\title{Supporting information to: Resolution Enhancement in Protein NMR Spectra by Deconvolution with Compressed Sensing Reconstruction}

\author[1]{Krzysztof Kazimierczuk}
\author[1,2]{Pawe\l~Kasprzak}
\author[3]{Panagiota S. Georgoulia}
\author[4,5]{Irena Mate\v{c}ko-Burmann}
\author[3,5]{ Bj\"orn M. Burmann}
\author[3]{Linn\'ea Isaksson}
\author[6]{Emil Gustavsson}
\author[3]{Sebastian Westenhoff}
\author[3,6]{Vladislav Yu. Orekhov}
\date{}                     %% if you don't need date to 

\affil[1]{\textit{Centre of New Technologies, University of Warsaw, Banacha 2C, 02-097 Warsaw, Poland }}

\affil[2]{\textit{Faculty of Physics, University of Warsaw, Pasteura 5, 02-093 Warsaw, Poland. }}

\affil[3]{\textit{Department of Chemistry and Molecular Biology, University of Gothenburg, 405 30 Gothenburg, Sweden}}

\affil[4]{\textit{Department of Psychiatry and Neurochemistry, University of Gothenburg, Gothenburg 405 30, Sweden}}

\affil[5]{\textit{Wallenberg Centre for Molecular and Translational Medicine, University of Gothenburg, Gothenburg 405 30, Sweden}}

\affil[6]{\textit{Swedish NMR center, University of Gothenburg, Box 465, 405 30, Gothenburg, Sweden}}
%            $^3$ Department of Mathematical Methods in Physics, Faculty of Physics, University of Warsaw, Pasteura 5, Warsaw, Poland\\

% \keywords{non-uniform sampling; temperature coefficients; osteopontin; intrinsically disordered proteins; HNCO}
\maketitle
%\subjclass[]{}
%\title[]{Supporting information to: Resolution Enhancement in Protein NMR Spectra by Deconvolution with Compressed Sensing Reconstruction}
% \keywords{ }
 
%\author{Krzysztof Kazimierczuk}
%\address{{\textit{Centre of New Technologies, University of Warsaw, Banacha 2C, 02-097 Warsaw, Poland }}}
%\author[1]{Paweł Kasprzak\fnref{fn1}}
%\author[2]{Panagiota S. Georgoulia}
%\author[3]{Irena Burmann}
%\author[2,4]{Vladislav Yu. Orekhov\corref{cor1} %\fnref{fn2}}

%\address[1]{Department of Chemistry and Molecular Biology, University of Gothenburg, 41390 Gothenburg, V\"{a}stra G\"{o}taland, Sweden}
%\address[2]{Department of Chemistry and Molecular Biology, University of Gothenburg, 41390 Gothenburg, Sweden}
%\address[3]{Department of Psychiatry and Neurochemistry at Institute of Neuroscience and Physiology, Blå Stråket 15, Sahlgrenska University, Gothenburg, 413 45 Gothenburg, Sweden}
%\address[4]{Swedish NMR center, Department of Chemistry and Molecular Biology, University of Gothenburg, 41390 Gothenburg, Sweden}
 
%\title{Signaling mechanism of phytochromes in solution}
\section{Deconvolution with Iteratively Re-weighted Least Squares}
Let  $f$ be the the subsampled  NMR (complex) signal measured at  $\{t_{1},\ldots, t_{k}\}$ and viewed   as the column vector $f\in\mathbb{C}^k$. The sampling schedule   $\{t_{1},\ldots, t_{k}\}$ is a fixed subset of the full sampling grid $\{l\Delta t:l=0,1,\ldots,n-1\}$. The CS methodology when applied in NMR context provides the  algorithms and theory  which enable for the recovery of the NMR spectrum $s\in\mathbb{C}^n$ from the sub-sampled signal, even if  $k\ll n$. 
The CS problem is formulated in terms of the 
%and describe dedicated algorithm  known as Iteratively Reweighted Least Squares (IRLS) \cite{ChartrandYin,Kazimierczuk20115556,KazimierczukCSConvexVsNonconvex}  
measurement matrix $F\in M_{k\times n}(\mathbb{C})$. In the NMR application of CS the matrix $F$ consists of  rows of the $n\times n$ inverse Fourier matrix   corresponding to the sampling schedule and we have \begin{equation} \label{e1} f = Fs. \end{equation} Note that Eq. \eqref{e1} with  known $f$ and unknown $s$ and $k<n$, has infinitely many solutions. The fundamental insight of the CS theory \cite{FoucartRauhutBook} specifies the NMR spectrum $s$ as the unique solution of the convex optimization problem   
\begin{equation}\label{e2}s = \argmin_{x\in\mathbb{C}^n}\left(\|Fx- f\|^2_2+\lambda\|x\|_1\right).\end{equation}
The first term in this sum promotes the consistency of tested $x$ with the measured data whereas, the second term promotes the sparseness of $x$ and $\lambda>0$ fixes the balance between the two. 

\textbf{The Eq. \eqref{e2} can be modified to take better account of the measurement noise. We can consider two cases:}
\begin{itemize}
    \item Unstructured noise: the covariance noise matrix $\Sigma$ is diagonal and isotropic, $\Sigma = \sigma^2\I$ where $\sigma$ is the standard deviation of the noise.  In order to take into an account  the level of  noise in the reconstruction framework \eqref{e2} one  sets $\lambda$  proportional to $\sigma^2$.
    \item Structured noise:  the covariance noise matrix  $\Sigma$ is not necessarily diagonal and isotropic.   The $\ell_2$ norm used in the data consistency term should be replaced by its weighted version $\|Fx- f\|^2_Q$, where  $Q=\Sigma^{-1}$ and  for a given  complex positive definite matrix $G$ and a complex vector $y$ we define $\|y\|^2_G = y^\dagger Gy$ where $y^\dagger$ is the conjugated transpose of $y$. In order to justify the proposed modification of the data consistency term let us discuss an instructive example. For that matter consider two  independent measurements $f_1 = f(t_1), f_2= f(t_2)$ of the signal in which the noise level of $f(t_1)$ is two times smaller then that of $f(t_2)$. The corresponding covariance matrix of $\Sigma$ is of the form \[\Sigma = \begin{bmatrix}
    \sigma^2&0\\0&4\sigma^2
    \end{bmatrix}\] where  $\sigma$ is   the noise level  entering $f(t_1)$. Let us denote $f^\#= Fx$  the signal corresponding to the spectrum $x$. The consistency term
    \[\|Fx - f\|_2^2 = |f^\#_1-f_1|^2+|f^\#_2-f_2|^2\] 
    entering the standard formulation of the CS-problem is replaced by 
    \[\|Fx - f\|_Q^2 =  (f^{\#\dagger} - f^{\dagger})\Sigma^{-1}(f^{\#} - f) = \frac{1}{ \sigma^2}|f^\#_1-f_1|^2+\frac{1}{ 4\sigma^2}|f^\#_2-f_2|^2
     \] Our  modification introduces the  weights in the data consistency term that correctly reflect the noise level of the corresponding measurements. Points with larger noise enter the sum with smaller weights. The above discussion and conclusion  easily generalizes to larger number of independent measurements. In case of the non-zero correlations between the noise of the measurements, $\Sigma$ must be first diagonalized in an appropriate  orthonormal basis and the above justification can be then repeated. 
\end{itemize}    Let us  note that for the unstructured noise   ($\Sigma =\sigma^2\I$)  we have  \begin{equation}\label{et2}\|Fx-f\|^2_Q = \sigma^{-2}\|Fx-f\|_2^2\end{equation} and thus the cases of structured and unstructured noise are consistent: \begin{align*}\argmin_{x\in\mathbb{C}^n}\left(\|Fx- f\|^2_Q+\lambda\|x\|_1\right)&=\argmin_{x\in\mathbb{C}^n}\left(\sigma^{-2}\|Fx- f\|_2^2+\lambda\|x\|_1\right)\\&=\argmin_{x\in\mathbb{C}^n}\left(\|Fx- f\|_2^2+\sigma^{2}\lambda\|x\|_1\right). 
\end{align*}  Let us also note that in each case  we  can absorb   $\lambda$   into $\Sigma$ by the possible scaling $\Sigma\rightsquigarrow \lambda\Sigma$. 

Now, we will  describe algorithm  known as Iteratively Reweighted Least Squares (IRLS) \cite{ChartrandYin,Kazimierczuk20115556,KazimierczukCSConvexVsNonconvex} dedicated for the solution of 
\begin{equation}\label{e4}s = \argmin_{x\in\mathbb{C}^n}\left(\|Fx- f\|^2_Q+\|x\|_1\right).\end{equation}
Note first, that the $\ell_1$ norm $\|x\|_1 = \sum_{i=1}^n|x_i|$ can be well approximated  by the weighted $\ell_2$ -norm 
$\sum_{i=1}^n|w_ix_i|^2$, where the weights $w_i=|x_i|^{-1/2}$ are regularized for very small $x_i$'s. This seemingly trivial observation  is the starting point of IRLS\cite{EmmanuelJ.Candesa}, which is an iterative procedure that solves  the quadratic problem 
\begin{equation}\label{e3}s_l = \argmin_{x\in\mathbb{C}^n}\|Fx- f\|^2_Q+\|w_lx\|_2^2\end{equation}
where $l$ is the iteration loop number  and $w_l$ is the weight vector corresponding to $s_{l-1}$ as described above. Thus defining $ D_l=\textrm{diag}(d_1,\ldots d_n)$ where $d_j = \frac{1}{|s_{l-1,j}|+\varepsilon}$   (we write $s_{l-1,j}$ for the $j$th component of the vector $s_{l-1}$), Eq.\eqref{e3} can be written in the form of generalized Tikhonov regularization 
\begin{equation}\label{e5}s_l = \argmin_{x\in\mathbb{C}^n}\|Fx- f\|^2_Q+\|x\|_{D_l}^2\end{equation}
and the latter can be solved explicitly
\[s_l = (F^*QF+D_l)^{-1}F^*Qf\]

\iffalse
\begin{equation}\label{wdef0}
\begin{aligned}s_{l} &= D_{l-1}^{-1}F^*(FD_{l-1}^{-1}F^*+\lambda)^{-1}f\\
  D_l&=\textrm{diag}(d_1,\ldots d_n)
\end{aligned}
 \end{equation}
where $d_j = \frac{1}{|s_{l-1,j}|^{1+l\delta}+\varepsilon}$
  (we write $s_{l-1,j}$ for the $j$th component of the vector $s_{l-1}$).
\fi

\subsection*{J-coupling in IRLS}  
 The J-modulation of the NMR signal $f$  is represented in vector language by a complex vector  
 $M\in\CC^k$.  For example $M$ corresponding to the $J$-coupling considered in the main text (i.e. approximately the same for all components) is of the form $M(t_j) = \cos(\pi Jt_j)$. The unmodulated version  $\tilde{f}$ of $f$  is defined by the equality $f = M \tilde f$, or more precisely  $ f(t_i)=M(t_i) \tilde f(t_i)$ where  $i\in\{1,\ldots, k\}$. Let  $C\in M_{k\times k}(\CC)$ denote the modulation (diagonal) matrix
 \[C = \textrm{diag}(M(t_1),\ldots,M(t_k)) \] The relation between  $\tilde f$, $f$ and the spectrum $\tilde s$  corresponding to  unmodulated signal $\tilde f$ is of the form 
 \[f = C\tilde{f} = CF\tilde s.\]
 Note, that while the noise in the measured signal $f$ is unstructured, the noise of $\tilde{f}$ is structured (since  $\tilde{f} = C^{-1}f$).  
 
  Denoting the spectrum related to modulated signal $f$ by  $  s$ we have $f = F s$. Remarkably, the spectrum $\tilde s$ is sparser than $s$ for typical modulations encountered in NMR. For example in the simplest case of  one dimensional J-modulated signal, singlets in $\tilde{s}$ are doubled in $s$, i.e. the number $\tilde m$ of significant elements of the spectrum  $\tilde{s}$  is approximately half of the number $m$  of the significant component of $s$. The standard estimation for the number of measurements $k$ required for the exact CS reconstruction is\cite{CandRomTao}
\begin{equation}\label{csest}k\sim m\log(n)\end{equation}  \textbf{This estimation and the preceding remark show that that  number of measurements $\tilde k$  required for the recovery of $\tilde{s}$  gets divided by the factor 2 when compared with  $k$  required for the recovery of $s$.}  The above observations motivate the development of a CS methodology dedicated to the signal in the  presence of modulations, which we formulate in what follows. 

Assuming that the J-modulated (measured) signal $f$ is  corrupted by the unstructured   noise with  covariance matrix $\Sigma$, the de-modulated signal  $\tilde f = C^{-1} f$ is corrupted by the structured  noise  with the  covariance matrix $  \tilde\Sigma = C^{-1} \Sigma C^{\dagger-1}$. In particular the weighting matrix $\tilde Q = \tilde\Sigma^{-1}$   is equal to $C^\dagger Q C$ and we get  
\iffalse This form of the noise should be taken into account in IRLS when computing the data consistency $\ell_2$ term by introducing the weights reflecting the relative noise strength of the given coordinate. This concept can be implemented  by replacing   $\|Fx-f\|_2^2$ in \eqref{e2} with its weighted version $\|Fx-f\|^2_Q$ where: $Q= \sigma^{-2}C^*C$ (c.f. the weighted  Tikhonov regularization) and 
  $\|b\|_Q^2 = b^*Qb$ 
for every vector $b$. Summarizing, the data consistency term is computed by the weighted $\ell_2$ norm where the weights are proportional to the modulation  factor. 
\fi

\begin{align*}\|Fx-\tilde f\|_{\tilde Q}^2 &= (Fx- \tilde f)^\dagger   C^\dagger Q C(Fx-f)\\&= (CFx- C\tilde f)^\dagger Q(CFx-C\tilde f)
\\&= (CFx-  f)^\dagger  Q  (CFx-  f)\\&=  \|CFx -  f\|_{Q}^2.\end{align*} 
%Note that in the standard NMR measurement setup the covariance matrix $\tilde\Sigma$ is a diagonal isotropic matrix with $\sigma^2$ on the diagonal and then $\|CFx - \tilde{f}\|_{\tilde{Q}}^2 = \sigma^{-2}\|CFx - \tilde{f}\|_{2}^2$. 
This computation  shows that the solution of    minimization problem 
\[\tilde{s} = \argmin_{x\in\mathbb{C}^n}\left(\|Fx- \tilde f\|^2_{\tilde Q}+\|x\|_1\right)\] coincides with that corresponding to 
\[\tilde{s} = \argmin_{x\in\mathbb{C}^n}\left(\|CFx-   f\|^2_{  Q}+\|x\|_1\right)\] and the latter can be found by the IRLS in the iterative procedure 
\begin{equation}\label{e6}\tilde s_l = \argmin_{x\in\mathbb{C}^n}\left(\|CFx- f\|^2_{Q}+\|x\|_{D_l}^2\right). \end{equation}   In the resulting spectrum the multiplets reflecting the modulation  (e.g. doublets) will be replaced by singlets. Keeping in mind the fact that in the standard experimental setup the measurement noise of  $f$   is described by  $ \Sigma = \sigma^2\I$, and substituting $\lambda = \sigma^2$ as explained above, we observe that Eq.\eqref{e6} boils down to the iterative procedure based on generalized Tikhonov regularization 
\begin{equation}\label{e7}\tilde s_l = \argmin_{x\in\mathbb{C}^n}\left(\|CFx- f\|^2_{2}+\lambda\|x\|_{D_l}^2\right). \end{equation} Remarkably this is the IRLS algorithm  in the orthodox form  (c.f. \eqref{e2}) applied to \begin{equation}\label{e8}\tilde s = \argmin_{x\in\mathbb{C}^n}\left(\|CFx-  f\|^2_2+\lambda\|x\|_1\right).\end{equation}
which in turn may be viewed  as mathematical formulation of the problem of CS-type of finding a sparse spectrum $\tilde s$ whose  consistency  with the measurement vector $f$ is given by the measurement matrix $CF$: $f = CF\tilde s$. 

Summarizing, in  the above considerations we explained the following two aspects:
\begin{itemize}
    \item we formulate the CS-problem  suitable for  the case of noisy measurements described by the  structured measurements noise; we also  describe the  solution of this problem and relate it with the Tikhonov regularization;
    \item we applied the modified CS-methodology to the J-modulated signals in NMR and explain the advantages of our approach when compared to the orthodox version of CS.  
\end{itemize}
\newpage
\FloatBarrier
\section{Protein samples and NMR spectroscopy}
A [U-\ce{^2H},\ce{^{15}N},\ce{^{13}C}] labeled sample of the monomeric photosensory module (57 kDa) $Dr$BphP$_{PSM}$ from \textit{Deinococcus radiodurans} was produced exactly as described in our previous study\cite{Gustavsson2020,BMRB27783}. The monomeric variant contains three mutations, which disrupt the dimer interface: F145S, L311E, and L314E\cite{Takala}.
%%\textbf{Add this citation here: Takala, Björling, et al, JBC, 2015: https://www.jbc.org/content/290/26/16383.full}   

\begin{figure}
    \centering
    \includegraphics[width=0.99\linewidth]{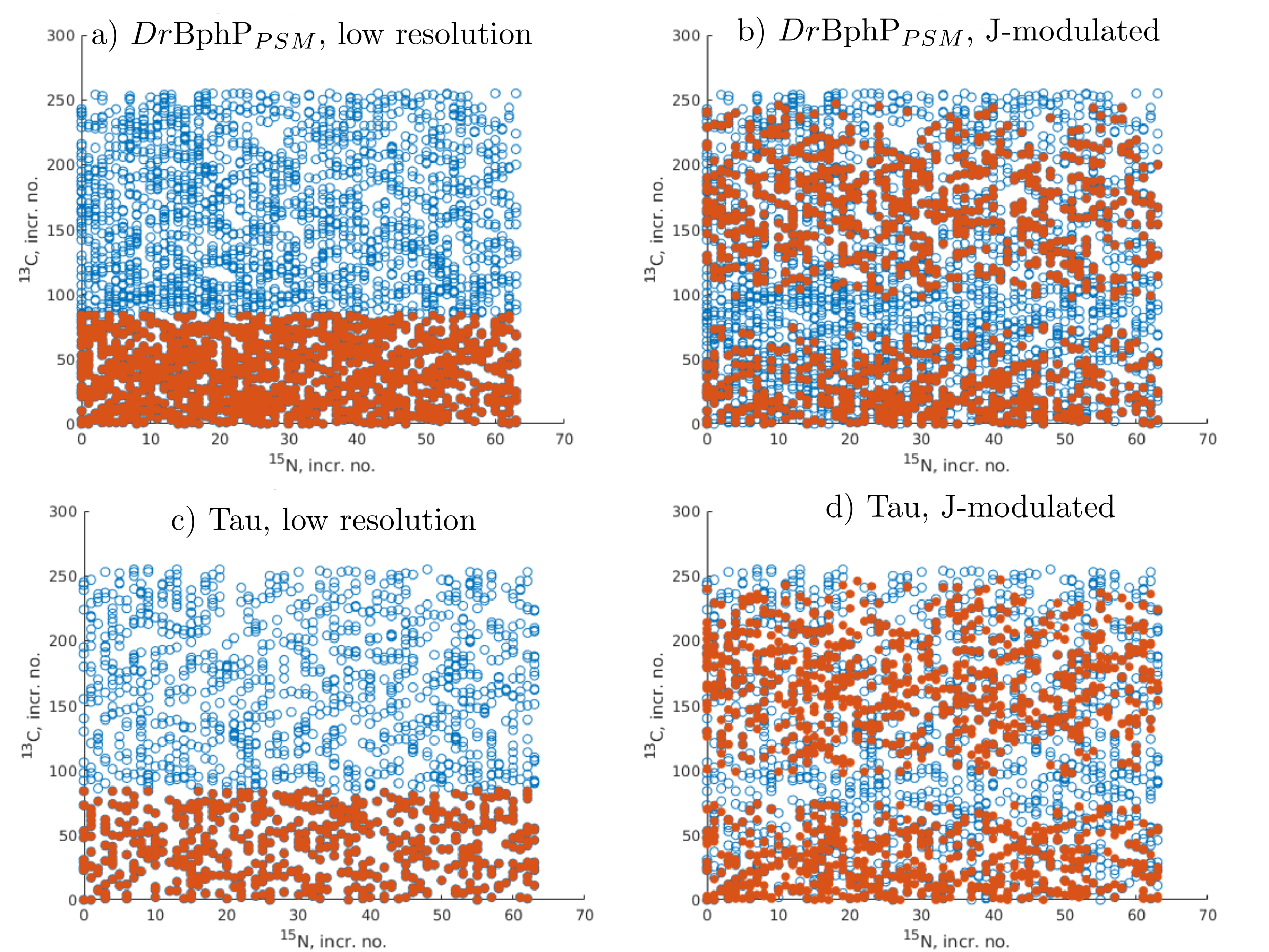}
    \caption{Sampling schedules used in the original experiments (blue) and sub-sampled data (red). The original schedules include 2930 points for $Dr$BphP$_{PSM}$ and 1552 for Tau and follow the relaxation-matched NUS scheme with $\textrm T_2=70$ ms for Tau (c) and d) and $\textrm T_2=200$ ms for $Dr$BphP$_{PSM}$ (c) and d). Schedules used for calculation of the "low-resolution" spectra, shown in panels a) and c) for $Dr$BphP$_{PSM}$ and Tau, respectively, were created by truncating the \ce{^{13}C} dimension to 14 ms which resulted in 1170 (a) and 555 (c) points.  The "J-modulated" schedules shown in panels b) and d), for $Dr$BphP$_{PSM}$ and Tau, respectively, were created by selecting 1200 points that matched the $|\cos(\pi t/J)|$ envelope in the  \ce{^{13}C} dimension (with elimination of points for which $|\cos(\pi t/J)|<0.2$).
   }
    \label{fig:Sampling_schedules}
\end{figure}

A 3D BEST-TROSY-HNCA experiment \cite{Solyom2013} was recorded for the $Dr$BphP$_{PSM}$ sample during 38.5 hours with 2930 relaxation-matched  NUS points (assumed sampling density decay rate $\textrm T_2=70$ ms) in the \ce{^{13}C^{\alpha}} dimension. The spectral widths (acquisition times) were 12.8 kHz (80 ms), 2.9 kHz (22 ms), and 6.0 kHz (42.4 ms), for \ce{^1H}, \ce{^{15}N}, and \ce{^{13}C} spectral dimensions, respectively. For the processing, the original NUS data set was sub-sampled to create the following data sets: 
\begin{itemize}
    \item The "traditional" low resolution spectrum processed using IRLS algorithm without the virtual decoupling, 1170 NUS points (15.4 hours of measurement time) were retained from the original NUS data with maximal evolution time of 14 ms for the \ce{^{13}C^{\alpha}} dimension (Figure \ref{fig:Sampling_schedules}a)). 
    \item The high resolution spectra processed using IRLS (undecoupled) and the region-selective D-IRLS algorithm (decoupled). The original NUS data was sub-sampled down to 1200 NUS points (15.8 hours of measurement time). The resulting sampling probability distribution for this spectrum corresponded to the sampling in the \ce{^{13}C} dimension matched to the both relaxation and J-coupling ($\textrm T_2=70$ ms, J=35 Hz), i.e. proportional to  $\exp(-t/{\textrm T_2})|\cos(\pi t/J)|$, with additional elimination of the points whose values of $|\cos(\pi t/J)|$ were less than 0.2 (Figure \ref{fig:Sampling_schedules}b)). 
\end{itemize}

A [U-\ce{^2H},\ce{^{15}N},\ce{^{13}C}] labeled sample of the longest human Tau protein isoform hTau40 with 441 residues was prepared as following: full length hTau40 with an amino-terminal His\textsubscript{6}-SUMO-Tag (in a modified pET28b plasmid, Genescript) was expressed in \textit{E. coli} BL21( $\lambda$DE3) Star\texttrademark (Novagen) cells. [U-\ce{^2H},\ce{^{15}N},\ce{^{13}C}] isotope (Merck) enriched protein was produced using 2xM9 minimal medium\cite{azatian2019increasing} supplemented with \ce{^{15}NH4Cl} and D-(\ce{^2H}/\ce{^{13}C})- glucose as the sole nitrogen and carbon sources, respectively, in \ce{D2O}. The cells were grown at 37 \textdegree C until an OD\textsubscript{600} $\approx$ 0.8. Expression was induced by addition of 1mM isopropyl-thiogalactoside (IPTG) (Thermo Scientific) for 16h at 22 \textdegree C. Cells were harvested by centrifugation and subsequently resuspended in lysis buffer (20mM NaPi, 500mM NaCl, pH~7.8), and lysed by an Emulsiflex C3 (Avestin) homogenizer. Cleared lysate was purified with HisTrap HP column (GE Healthcare). Fractions containing hTau40 were pooled and dialyzed over-night, against human SenP1 cleavage buffer (20 mM TrisHCl, 150 mM NaCl, 1 mM DTT, pH~7.8). After dialysis SenP1 protease (Addgene \#16356)\cite{mikolajczyk2007small} was added and the enzymatic cleavage was performed for 4 hours at room temperature, followed by a second HisTrap HP column step. Fractions containing cleaved hTau40 in the flow-through were collected, concentrated, and subsequently purified by gel filtration using a HiLoad 10/60 200 pg (GE Healthcare) pre-equilibrated with NMR buffer (25mM NaPi, 50mM NaCl, 1mM EDTA, pH~6.9). The pure hTau40 fractions were concentrated to about 500 $\mu$M, flash frozen in liquid nitrogen, and stored at -80 \textdegree C till usage.

A 3D BEST-TROSY-HN\ce{^{13}C^{\alpha}} experiment \cite{Solyom2013} for Tau40 was recorded and processed similar to the $Dr$BphP$_{PSM}$ spectrum described above. Namely, it was acquired during 13 hours with 1552 NUS points using spectral width (acquisition times) of 9.6 kHz (106 ms), 2.9 kHz (22 ms), and 6.0 kHz (42.4 ms), for \ce{^1H}, \ce{^{15}N}, and \ce{^{13}C} spectral dimensions, respectively. For the processing, the originally random NUS data set was sub-sampled to create the following data sets:
\begin{itemize}
\item The "traditional" low resolution spectrum was processed using IRLS algorithm without the virtual decoupling, 555 NUS points (4.6 hours of measurement time) were retained from the original NUS data with a maximal evolution time of 14 ms for the \ce{^{13}C^{\alpha}} dimension (Figure \ref{fig:Sampling_schedules}c)).  
\item The high resolution spectra processed using IRLS (undecoupled) and the region-selective D-IRLS algorithm (decoupled). The original NUS data was sub-sampled down to 1200 NUS points (10 hours of measurement time). The resulting sampling probability distribution for this spectrum corresponded to the sampling in the \ce{^{13}C^{\alpha}} dimension matched to the both relaxation and J-coupling ($\textrm T_2=200$ ms, J=35 Hz), i.e. proportional to  $\exp(-t/{\textrm T_2})\cos(\pi t/J)$, with additional elimination of the points whose values of $|\cos(\pi t/J)|$ were less than 0.2 (Figure \ref{fig:Sampling_schedules}d)). 
\end{itemize}
\newpage
% \FloatBarrier
% \section{Spectra of proteins TAU and $Dr$BphP$_{PSM}$}
\begin{figure}
    \centering
    \includegraphics[width=0.9\linewidth]{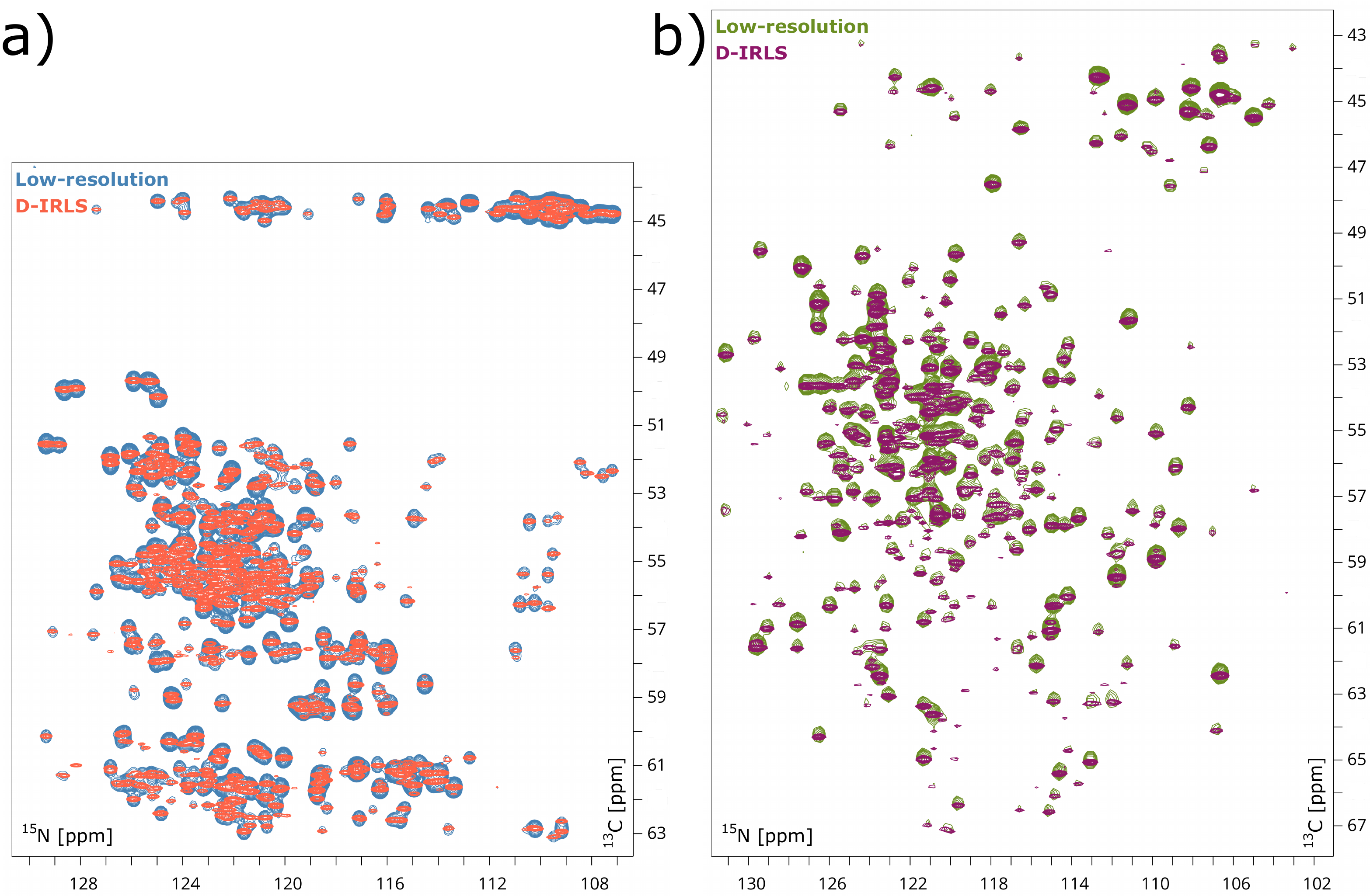}
    \caption{\ce{^{15}N}/\ce{^{13}C} projections of 3D HNCA spectra of a) Tau and b) $Dr$BphP$_{PSM}$ proteins. The plots show superimposed traditional "low-resolution" and high-resolution deconvoluted spectra with 14~ms and 42.4~ms of maximum evolution time in the \ce{^{13}C} dimension, respectively.}
    \label{fig:tau_CNproj}
\end{figure}
\FloatBarrier
\newpage
\section{Simulations with synthetic peaks}
\begin{figure}[h]
    \centering
    \includegraphics[width=\linewidth]{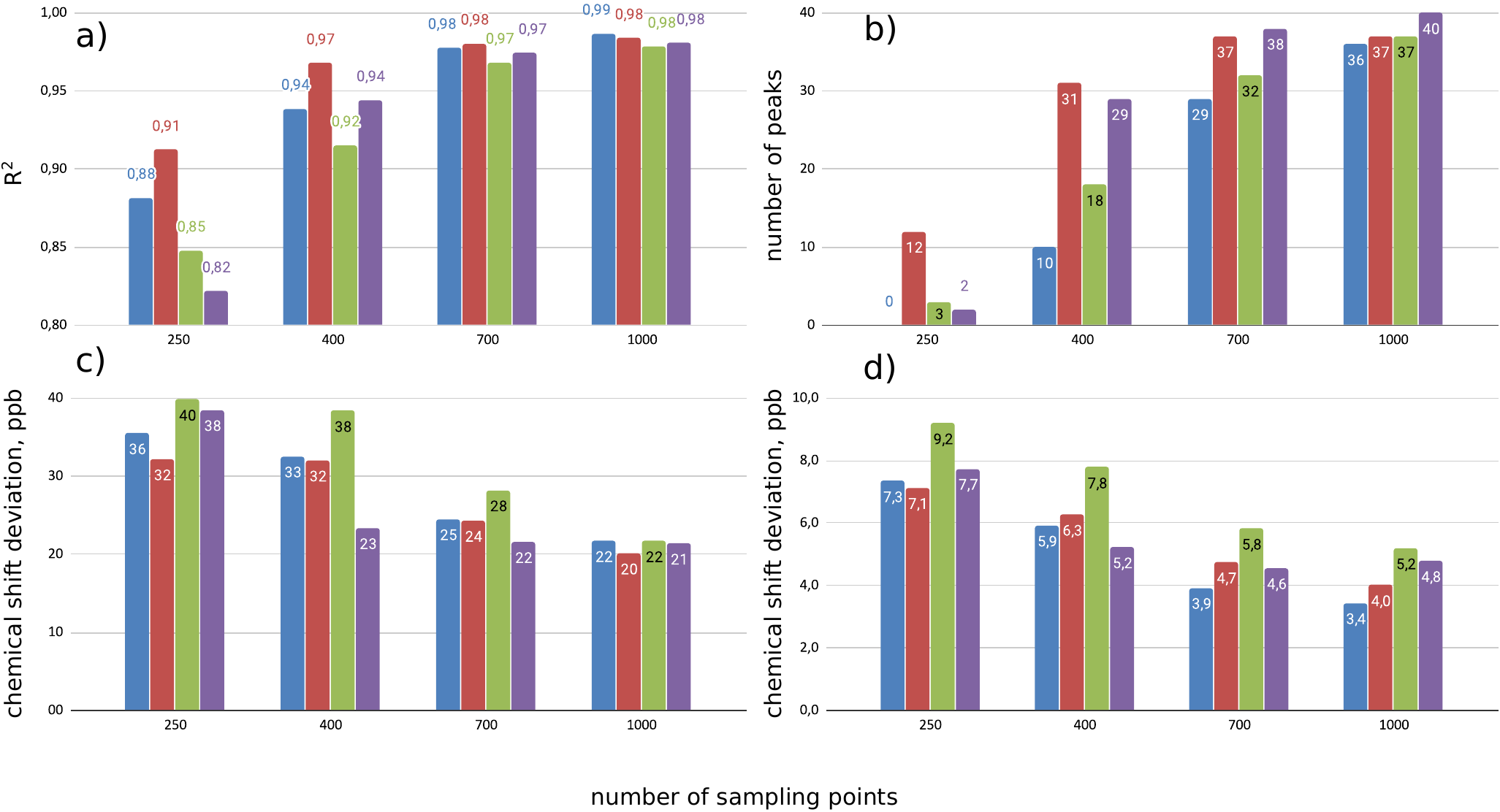}
    \caption{The results of simulations with synthetic peaks injected into the experimental spectrum of Tau protein. The columns correspond to: undecoupled IRLS (blue), D-IRLS with (red) and without (green) Gly-region extraction, all  with J-modulated sampling scheme, and D-IRLS with Gly-region extraction with non-modulated sampling (purple). The plots show: a) $R^2$ coefficients of peak intensity reconstruction, b) number of detected weak peaks out of 42 lowest intensity injected peaks, c) deviation of peak position in \ce{^{15}N} dimension, d) deviation of peak position in \ce{^{13}C^{\alpha}} dimension. Numbers at the tops of the columns indicate their exact heights.}
    \label{fig:correlation_plots}
\end{figure}

Synthetic peaks were injected into the spectrum of Tau protein in order to estimate accuracy of the spectra reconstructions. Comparison was performed between the reconstruction obtained using two types of NUS sampling schedules:
\begin{itemize}
    \item matched to both \ce{^{13}C^{\alpha}} transverse relaxation and J-modulation
    \item matched to the \ce{^{13}C^{\alpha}} relaxation only
\end{itemize}
    and four calculations modes: 
\begin{itemize}    
    \item traditional low resolution spectrum with \ce{^{13}C^{\alpha}} maximum evolution time of 14 ms reconstructed using IRLS algorithm without deconvolution of the J-coupling
    \item high resolution spectrum with \ce{^{13}C^{\alpha}} maximum evolution time of 42 ms - IRLS reconstruction without the deconvolution
    \item high resolution spectrum with \ce{^{13}C^{\alpha}} maximum evolution time of 42 ms - IRLS reconstruction with the deconvolution (D-IRLS) for the whole spectrum 
    \item region-selective processing with IRLS for the Gly region ($<$45 ppm in \ce{^{13}C^{\alpha}}) and D-IRLS for the rest of the spectrum (as in Figure 1 in the main text).
\end{itemize}     
    Each version of the reconstruction was calculated 15 times using selected with a corresponding random distribution (sub-sampled) NUS data sets from the larger pool of measured data (total 1552 flat-random NUS). In each calculation 20 peaks, including 5 Gly-type peaks, were injected with random positions (without overlap between each other and the existing Tau protein peaks) and intensities varying in the range 0.05-1.0 of a typical medium strong peak in the original Tau spectrum. The non-Gly signals were injected as the doublets with random J-coupling value in the range 35$\pm$5 Hz.  The resulting peaks  were automatically picked in the reconstructed spectra using function pkFindROI in nmrPipe software\cite{Delaglio1995}.  Thus, in each of the eight variants of the spectrum reconstructions, up to 225 non-Gly synthetic peaks were picked and quantified. Statistics on the peaks in accuracy of the peak intensities and positions is shown in Figure \ref{fig:correlation_plots}, where the comparisons are given for four calculations using 250, 400, 700, and 1000 NUS points.    

\FloatBarrier

\FloatBarrier

\newpage
\bibliography{main}

\begin{thebibliography}{10}

\bibitem{FoucartRauhutBook}
Simon Foucart and Holger Rauhut.
\newblock {\em {A Mathematical Introduction to Compressive Sensing}}.
\newblock Wiley, 2010.

\bibitem{ChartrandYin}
Rick Chartrand and Wotao Yin.
\newblock {Iteratively reweighted algorithms for compressive sensing}.
\newblock In {\em ICASSP}, pages 3869--3872. IEEE, 2008.

\bibitem{Kazimierczuk20115556}
Krzysztof Kazimierczuk and Vladislav~Yu. Orekhov.
\newblock {Accelerated NMR spectroscopy by using compressed sensing}.
\newblock {\em Angewandte Chemie - International Edition}, 50(24):5556--5559,
  2011.

\bibitem{KazimierczukCSConvexVsNonconvex}
Krzysztof Kazimierczuk and Vladislav~Yu. Orekhov.
\newblock {The comparison of convex and non-convex compressed sensing applied
  in multidimensional NMR}.
\newblock {\em Journal of Magnetic Resonance}, 223(0):1--10, 2012.

\bibitem{EmmanuelJ.Candesa}
Emmanuel~J. Cand{\`{e}}s, Michael~B. Wakin, and Stephen~P. Boyd.
\newblock {Enhancing sparsity by reweighted L1 minimization}.
\newblock {\em Journal of Fourier Analysis and Applications}, 14(5-6):877--905,
  2008.

\bibitem{CandRomTao}
E.~J. {Candes}, J.~{Romberg}, and T.~{Tao}.
\newblock Robust uncertainty principles: exact signal reconstruction from
  highly incomplete frequency information.
\newblock {\em IEEE Transactions on Information Theory}, 52(2):489--509, Feb
  2006.

\bibitem{Gustavsson2020}
Emil Gustavsson, Linn{\'{e}}a Isaksson, Cecilia Persson, Maxim Mayzel, Ulrika
  Brath, Lidija Vrhovac, Janne~A. Ihalainen, B.~G{\"{o}}ran Karlsson, Vladislav
  Orekhov, and Sebastian Westenhoff.
\newblock {Modulation of Structural Heterogeneity Controls Phytochrome
  Photoswitching}.
\newblock {\em Biophysical Journal}, 118(2):415--421, 2020.

\bibitem{BMRB27783}
{}bmrb entry 10.13018/bmr27783.

\bibitem{Takala}
H.~Takala, A.~Björling, M.~Linna, S.~Westenhoff, and J.A. Ihalainen.
\newblock {Light-induced Changes in the Dimerization Interface of
  Bacteriophytochrome}.
\newblock {\em The Journal of Biological Chemistry}, 290(26):16383--16392,
  2015.

\bibitem{Solyom2013}
Zsofia Solyom, Melanie Schwarten, Leonhard Geist, Robert Konrat, Dieter
  Willbold, and Bernhard Brutscher.
\newblock {BEST-TROSY experiments for time-efficient sequential resonance
  assignment of large disordered proteins}.
\newblock {\em Journal of Biomolecular NMR}, 55(4):311--321, 2013.

\bibitem{azatian2019increasing}
Stephan~B Azatian, Navneet Kaur, and Michael~P Latham.
\newblock Increasing the buffering capacity of minimal media leads to higher
  protein yield.
\newblock {\em Journal of Biomolecular NMR}, 73(1-2):11--17, 2019.

\bibitem{mikolajczyk2007small}
Jowita Mikolajczyk, Marcin Drag, Mikl{\'o}s B{\'e}k{\'e}s, John~T Cao, Ze'ev
  Ronai, and Guy~S Salvesen.
\newblock Small ubiquitin-related modifier (sumo)-specific proteases profiling
  the specificities and activities of human senps.
\newblock {\em Journal of Biological Chemistry}, 282(36):26217--26224, 2007.

\bibitem{Delaglio1995}
F.~Delaglio, S.~Grzesiek, G.~W. Vuister, G.~Zhu, J.~Pfeifer, and A.~Bax.
\newblock {NMRPipe: A multidimensional spectral processing system based on UNIX
  pipes}.
\newblock {\em Journal of Biomolecular NMR}, 6(3):277--293, 1995.

\end{thebibliography}
\bibliographystyle{unsrt}
\end{document}